\documentclass[twocolumn]{webofc}

\usepackage[varg]{txfonts}   

\usepackage{hyperref}
\usepackage[squaren,thinspace,mediumqspace,Gray]{SIunits}
\usepackage{xspace}

\hypersetup{
	colorlinks,
	linkcolor=blue,
	citecolor=cyan
}

\newcommand{\perkeo}{\textsc{Perkeo}\xspace}
\providecommand{\ep}{$e^{\scriptscriptstyle-}\!/\!p^{\scriptscriptstyle+}$\xspace}

\graphicspath{{principle/}{property/}{exp_prop/}}

\begin{document}

\title{Design of the Magnet System of the Neutron Decay Facility PERC}

\author{
\firstname{Xiangzun} \lastname{Wang}\inst{1,2} \and
\firstname{Carmen} \lastname{Ziener}\inst{3} \and
\firstname{Hartmut} \lastname{Abele}\inst{1} \and
\firstname{Stefan} \lastname{Bodmaier}\inst{2} \and
\firstname{Dirk} \lastname{Dubbers}\inst{3} \and
\firstname{Jaqueline} \lastname{Erhart}\inst{1} \and
\firstname{Alexander} \lastname{Hollering}\inst{2} \and
\firstname{Erwin} \lastname{Jericha}\inst{1} \and
\firstname{Jens} \lastname{Klenke}\inst{4} \and
\firstname{Harald} \lastname{Fillunger}\inst{1} \and
\firstname{Werner} \lastname{Heil}\inst{5} \and
\firstname{Christine} \lastname{Klauser}\inst{1,2,7} \and
\firstname{Gertrud} \lastname{Konrad}\inst{1,6} \and
\firstname{Max} \lastname{Lamparth}\inst{2,3} \and
\firstname{Thorsten} \lastname{Lauer}\inst{4} \and
\firstname{Michael} \lastname{Klopf}\inst{1} \and
\firstname{Reinhard} \lastname{Maix}\inst{1,2,3} \and
\firstname{Bastian} \lastname{Märkisch}\inst{2,3}\fnsep\thanks{\email{maerkisch@ph.tum.de}} \and
\firstname{Wilfried} \lastname{Mach}\inst{1} \and
\firstname{Holger} \lastname{Mest}\inst{3} \and
\firstname{Daniel} \lastname{Moser}\inst{6} \and
\firstname{Alexander} \lastname{Pethoukov}\inst{7} \and
\firstname{Lukas} \lastname{Raffelt}\inst{2,3} \and
\firstname{Nataliya} \lastname{Rebrova}\inst{3} \and
\firstname{Christoph} \lastname{Roick}\inst{2,3} \and
\firstname{Heiko} \lastname{Saul}\inst{1,2,4} \and
\firstname{Ulrich} \lastname{Schmidt}\inst{3} \and
\firstname{Torsten} \lastname{Soldner}\inst{7} \and
\firstname{Romain} \lastname{Virot}\inst{7} \and
\firstname{Oliver} \lastname{Zimmer}\inst{7}
\ (PERC Collaboration)
}

\institute{
Technische Universität Wien, Atominstitut, Stadionallee 2, 1020 Wien, Austria \and 
Physik-Department, Technische Universität München, James-Franck-Str. 1, 85748 Garching, Germany \and 
Physikalisches Institut, Universität Heidelberg, Im Neuenheimer Feld 226, 69120 Heidelberg, Germany \and
Forschungs Neutronenquelle Heinz Maier-Leibnitz, Technische Universität München, Lichtenbergstr. 1, 85748 Garching, Germany \and
Johannes-Gutenberg Universität, Staudingerweg 7, 55128 Mainz, Germany \and
Stefan-Meyer-Institut, Boltzmanngasse 3, 1090 Vienna, Austria \and
Institut Laue-Langevin, 71 avenue des Martyrs,
CS 20156, 38042 Grenoble Cedex 9, France
}


\abstract{
The PERC (Proton and Electron Radiation Channel) facility is currently under construction at the research reactor FRM~II, Garching. It will serve as an intense and clean source of electrons and protons from neutron beta decay for precision studies. It aims to contribute to the determination of the Cabibbo-Kobayashi-Maskawa quark-mixing element $V_{ud}$ from neutron decay data and to search for new physics via new effective couplings.

PERC's central component is a $\unit{12}{m}$ long superconducting magnet system. It hosts an $\unit{8}{m}$ long decay region in a uniform field. An additional high-field region selects the phase space of electrons and protons which can reach the detectors and largely improves systematic uncertainties. We discuss the design of the magnet system and the resulting properties of the magnetic field.
}


\maketitle

\section{Introduction} 

Free neutron $\beta$-decay provides a clean platform for precision searches for physics beyond the standard model \cite{Dub11,Gon19}. This is enabled by the absence of nuclear structure effects and small and well-known radiative corrections\footnote{We note that the common radiative corrections changed recently \cite{Sen18}.}. Within the standard model, free neutron decay is completely described by only three parameters which have to be determined experimentally. These are the ratio of axial- and vector- coupling constants $\lambda = g_\mathrm{A} / g _\mathrm{V}$, the element $V_{ud}$ of the Cabibbo-Kobayashi-Maskawa quark mixing matrix, and the Fermi coupling constant $G_\mathrm{F}$. The latter is known with very high precision from muon decay \cite{Web11}. $\lambda$ is most precisely determined from a measurement of the parity-violating beta asymmetry \cite{Mun13,Bro18,Mae18}. In combination with a measurement of the neutron lifetime $\tau_{n}$, the matrix element $V_{ud}$ can be determined. The precision reached is competitive with single measurements entering the most precise determination from superallowed nuclear decays and the results are in agreement \cite{Mae18,Dub19}. Within the framework of effective field theories \cite{Bha12, Cir13, Gon19}, neutron beta decay experiments test for deviations from the $V-A$ theory of the standard model, i.e. hypothetical scalar and tensor interactions, as well as right handed currents. A comparison of the experimentally determined $\lambda$ to lattice QCD results limits right handed couplings \cite{Cha18}.

Observables in neutron decay are the lifetime $\tau_n$, (angular) correlation coefficients (see \cite{Jac57, Dub11}), and the energy spectra of the decay protons and electrons. There are currently strong world-wide efforts to measure the beta asymmetry $A$ \cite{Dub08, Bro18, Mae18}, the neutrino asymmetry $B$ \cite{Dub08,Bro17}, the proton asymmetry $C$ \cite{Roi18b, Dub08}, the electron-neutrino correlation $a$ \cite{Dar17, Bae13, Kon09}, the Fierz interference term $b$ \cite{Hic17, Sau18, Bae13, Bro17, Dub08} and several coefficients related to the transverse electron polarisation \cite{Bod16}. A non-zero Fierz interference term would be of special interest as it would directly imply novel scalar or tensor interactions. More experiments are proposed at the upcoming European Spallation Source (ESS) using a new intense beam line for particle physics \cite{Sol18}.

In this manuscript we discuss the facility PERC (Proton and Electron Radiation Channel) \cite{Dub08,Kon12}, which is currently under construction at the new MEPHISTO beam site \cite{MEPHISTO} at the research reactor FRM~II, Garching, Germany. The aim of PERC is to provide a clean, bright and versatile source of neutron decay products (electrons and protons) to enable measurements which improve on the results of \perkeo~II \cite{Mun13} by at least an order of magnitude using specialised secondary spectrometers.

\begin{figure*}[tb]
	\centering
	\includegraphics[width=0.9\textwidth]{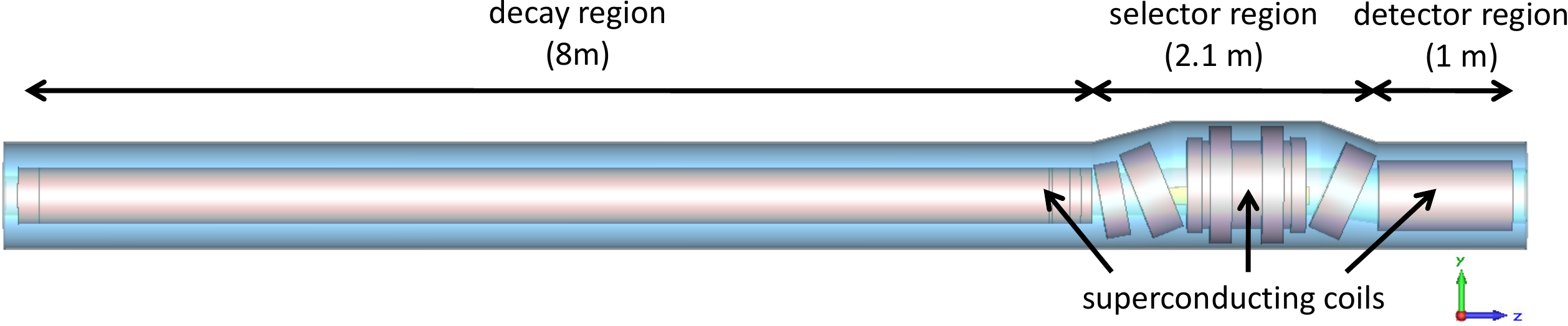}
	\caption{Sketch of the PERC magnet system. Three sections are identified: the long upstream solenoid contains the decay volume, the selector region contains the magnetic filter and separates charged decay products from the neutron beam, and the detector section contains detector systems or serves for transport to specialised spectrometers downstream.}
	\label{figure:perc sketch1}
\end{figure*}

The concept of PERC, as well as a study of systematic effects is described in detail in \cite{Dub08}. Its main features are a large active volume of $\unit{5 \times 5 \times 800}{cm^3}$ inside a neutron guide to observe the decay of free cold neutrons and a magnetic filter which limits the phase space of electrons and protons and strongly improves systematics. Together this enables distortion-free measurements on the level of $10^{-4}$ \cite{Dub08}.

Developments by the collaboration that are important for PERC include non-depolarising supermirror neutron guides \cite{Reb14}, precision neutron polarimetry \cite{Zimmer99,Klauser12,Klauser13}, electron and proton detection systems based on existing techniques \cite{Raf16, Roi18b} and novel concepts \cite{Wang13, Kon15}, calibration sources for proton spectroscopy \cite{Vir17} and calibration techniques for electron detection \cite{Roi18a}, refined data analysis and understanding of the detector response \cite{Roi18b, Sau18, Mae18, Roi19}, and the transport of particles in the magnetic field \cite{Dub14}.

In this paper we focus on the design of PERC's magnet system and give insight on its optimisation. An in depth discussion of the design decisions can be found in \cite{Wang13dis, Zie15} and the design was briefly introduced in \cite{Kon12}. 
The field optimisation was partially performed using the commercial package CST~Studio \cite{CST}, the RADIA \cite{Chu98} add-on to Wolfram Mathematica, as well as the \texttt{magfield3} code developed by F.~Glück \cite{Glu11}.

\section{The PERC magnet system}

The main component of PERC is a $\unit{12}{m}$ long superconducting magnet system.  The strong magnetic field defines a quantisation axis and guides the neutron spin in case of measurements with a polarised beam. Electrons and protons are guided by the field towards detector systems at the rear end and are separated from the cold neutron beam in a curved magnetic field section. 

PERC will be installed at the new MEPHISTO beamline for cold neutrons currently under construction at the FRM~II. The beamline will provide several optional features such as a wavelength selector, chopper, polariser and spin flipper to allow for different measurement modes. The projected neutron capture flux density is $\unit{2\cdot 10^{10}}{s^{-1} cm^{-2}}$.


The PERC magnet consists of three different sections as illustrated in Fig.~\ref{figure:perc sketch1}: the decay region, the selector region which contains the magnetic filter and separates charged decay products from the neutron beam, and the detector section which contains detector systems or serves to transport the electrons and protons to specialised spectrometers downstream.

The decay region consists of an $\unit{8}{m}$ long superconducting solenoid that features a homogeneous magnetic field of up to $B_0 = \unit{1.5}{T}$. A non-depolarising $m=2$ supermirror neutron guide is located inside the warm bore of the solenoid and contains the actual \emph{decay volume}. In order not to alter measurements of the beta and proton asymmetry significantly, this guide must preserve the neutron polarisation at the $10^{-4}$ level per bounce and will hence be based on copper and titanium \cite{Reb14}. While only a small fraction of the neutrons decay within the guide, the majority is absorbed by the neutron beamstop, which is located in the subsequent selector region. The charged decay products follow the magnetic field lines towards both ends of the solenoid.

\begin{figure}[!t]
	\centering
	\includegraphics[width=0.45\textwidth]{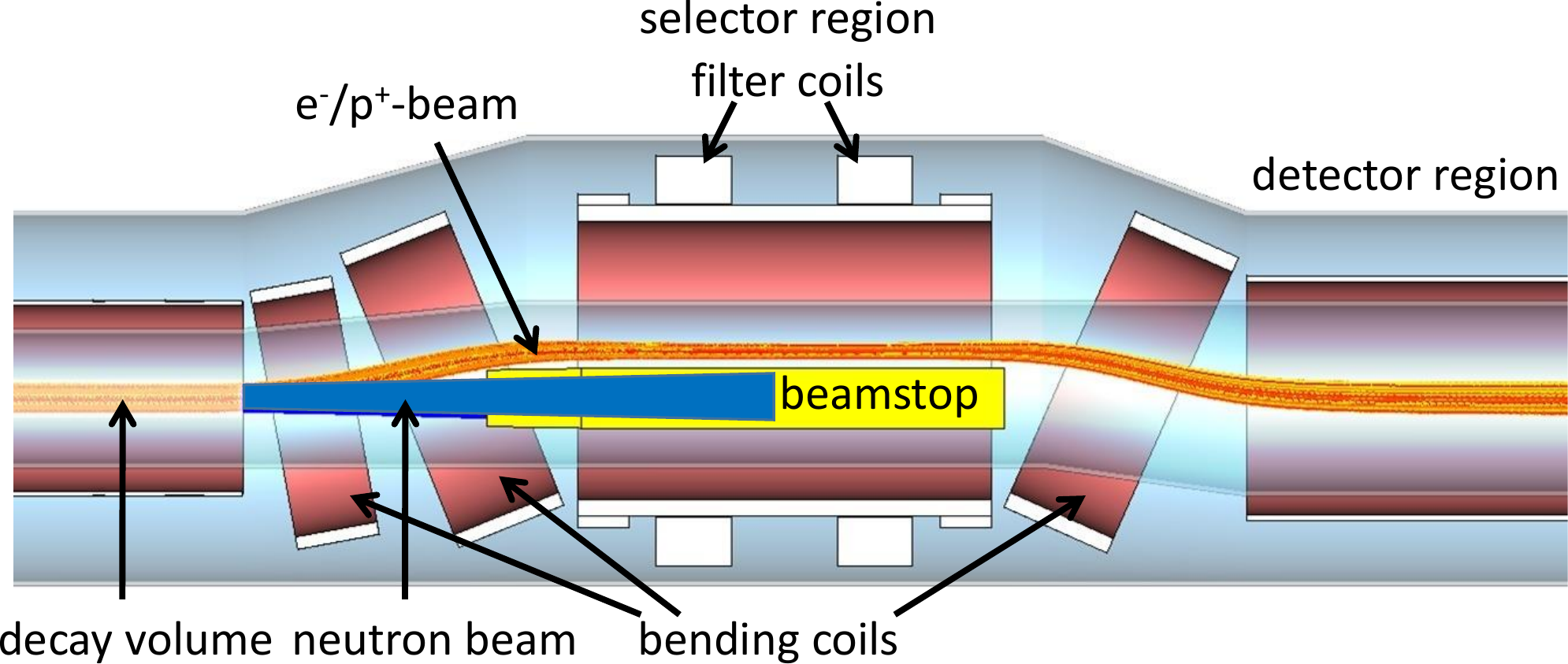}
	\caption{Close-up of the selector region: coils are drawn in red, simulated \ep-trajectories in orange. The remaining cold neutron beam (blue) coming from the left side is absorbed in the beamstop.}
	\label{figure:perc sketch2}
\end{figure}

In the selector region, three tilted coils create a curved magnetic field of maximum strength $B_1$. This decouples the \ep-beam from the neutron beam as illustrated in Fig.~\ref{figure:perc sketch2}. Behind the neutron beamstop, the \ep are guided back to the central axis. This is different to the original design described in \cite{Dub08}. It enables varying the $B_1$ field over a wide range without drastically affecting the particle trajectories and to implement an efficient ``two pinhole'' shielding scheme to suppress background created by the internal guide or neutron beam components upstream of PERC. Since the \ep-beam in the detector region is on-axis with the magnet system, it can be coupled to additional magnet systems downstream.

The field $B_1 = \unit{3\dots6}{T}$ in the selector region acts as a magnetic mirror. This allows to filter electrons and protons according to their angle of emission $\theta_0$ relative to the magnetic field direction, with the critical angle $\theta_c$ in the adiabatic limit only being dependent on the magnetic field:
\begin{equation}
	\theta_0 \leqslant \theta_c = \arcsin \sqrt{\frac{B_\mathrm{loc}}{B_1}}
\end{equation}
where $B_\mathrm{loc} \approx B_0$ is the magnetic field at the location of the decay. Hence $\theta_c \approx 30\degree$ for $B_0=\unit{1.5}{T}$ and $B_1=\unit{6}{T}$.

After the selector region, the field strength decreases to $B_2 = \unit{0.1\dots1}{T}$. For a typical value of $B_2 = \unit{0.5}{T}$, the maximum pitch angle of electrons and protons hence decreases to
\begin{equation}
	\theta_\mathrm{det} \leqslant \arcsin\sqrt{\frac{B_2}{B_1}}
	= 16.8\degree.
	\label{form:det theta}
\end{equation}
This reduces backscatter effects from the detector by a factor of $\approx 2$ compared to \perkeo.

Different detector systems specialised on measurements of certain observables will be attached here in order to detect electrons or protons, or both simultaneously. An example of such a spectrometer is the $\mathbf{R}\times\mathbf{B}$ momentum spectrometer NoMoS currently under development \cite{Wang13, Kon15}.

\begin{figure}[ht]
	\centering
	\includegraphics[]{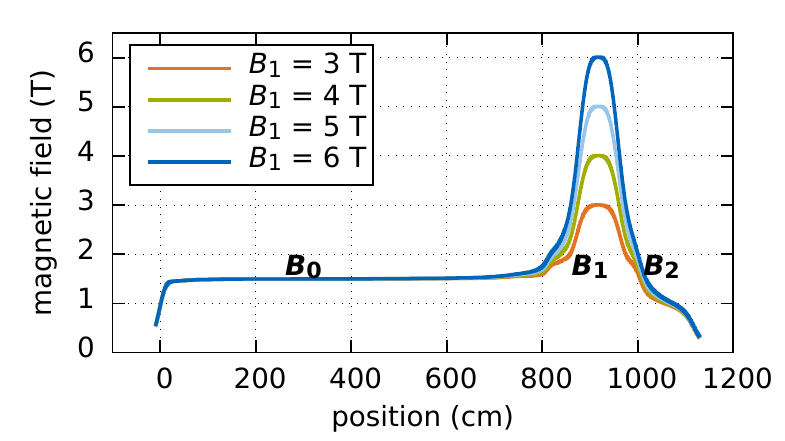}
	\caption{Magnetic field strength along the central field line, for different values of $B_1$ between $3$ and $\unit{6}{T}$, and $B_0=\unit{1.5}{T}$ and $B_2=\unit{0.5}{T}$ fixed.}
	\label{figure:perc field0}
\end{figure}

Fig.~\ref{figure:perc field0} shows the magnetic field strength along the central field line.  The effect of the magnetic filter can be varied by adjusting both $B_0$ and $B_1$ such that $2 < B_1 / B_0 < 12$ to either optimise for a specific measurement or to test the corresponding systematic effect. This extends the range envisaged in \cite{Dub08}.

\subsection*{Beta asymmetry}

An important quantity to be measured with PERC is the beta asymmetry parameter $A$ in neutron decay. This parity violating observable describes an asymmetry in the direction of the momentum $p_e$ of the decay electron with respect to the neutron spin:
\begin{equation}
  \mathrm{d}\Gamma_n\left(E_e\right) \propto 1 + A P_n \frac{p_e}{E_e} \cos\left(\theta_0\right),
\end{equation}
where $P_n$ is the neutron polarisation and $E_e$ is the electron energy.
With PERC, an experimental asymmetry will be determined from electron count rates for the two opposite directions of the neutron beam polarisation:
\begin{equation}
A_{\text{exp}}\left(E_e\right) = \frac{N^{\uparrow}\left(E_e\right) - N^{\downarrow}\left(E_e\right)}{N^{\uparrow}\left(E_e\right) + N^{\downarrow}\left(E_e\right)},
\end{equation}
where $\uparrow$ refers to the average neutron polarisation pointing to the detector and $\downarrow$ pointing away. This method has been successfully applied in previous measurements of the beta asymmetry, including the currently most precise determination of $\lambda$ \cite{Mae18}. However, in these measurements no magnetic filter has been applied. Taking into account the new PERC feature of phase space filtering, the relation between the experimental asymmetry and the beta asymmetry can be written as:
\begin{equation}
    A_\mathrm{exp}(E_e) = \frac{1}{2}\,A \,P_n\, \left( 1+\sqrt{1-\frac{B_0}{B_1}} \right)\frac{p_e}{E_e},
  \label{form:asymmetry perc}
\end{equation}
where we have neglected theoretical corrections. Note that the phase space filtering increases the measured asymmetry, while lowering the rate of detected events.

\section{Magnets and Field Properties}

The magnet system of PERC consists of $10$ different superconducting solenoids, not counting correction windings. The most complex magnet assembly is the selector which alone consists of $5$ solenoids, see Fig.~\ref{figure:perc sketch2}. 
Four individual power supplies allow to tune the fields in the long solenoid $B_0$, the selector region $B_1$ and the detector region $B_2$ separately. The selector region is driven by two power supplies: one creates the $\unit{3}{T}$ base field and the other drives the split pair configuration (filter coils) to increase the field up to $\unit{6}{T}$, see Fig.~\ref{figure:perc sketch2}.\footnote{The technical reason for the fourth power supply is the use of different types of superconducting wire} in the decay and selector regions.

\subsection{Decay region}

The $\unit{8}{m}$ \emph{long solenoid} generates the magnetic field $B_0$ in the decay volume. Additional windings or small gaps in the winding package at both ends of the solenoid improve the magnetic field uniformity.

\begin{figure}[!b]
	\centering
		\includegraphics[]{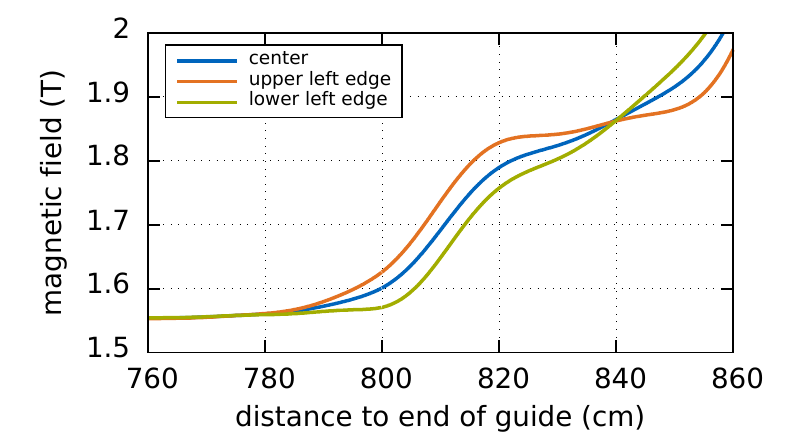}
	\caption{Magnetic field strength $B$ along the center and edges of the \ep-beam from the decay volume in the neutron guide to the selector for $B_1 = \unit{3}{T}$.}
	\label{figure:perc field bending}
\end{figure}

The magnetic field of PERC must not have local minima where charged particles may be trapped temporarily. Within the decay volume the resulting unpredictable losses and delays would hamper any high precision measurement. But even outside the neutron beam such traps could be filled by rest gas interaction leading to undesired background. Fig.~\ref{figure:perc field bending} shows the field strength along field lines originating in the center and at the edges of the neutron guide (with a cross section of $60 \times \unit{60}{mm^2}$) for the critical $B_1 = \unit{3}{T}$ configuration, i.e. with the minimum field change between the decay and selector regions. Field minima are absent.

\begin{figure}[!htb]
	\centering
	\includegraphics[]{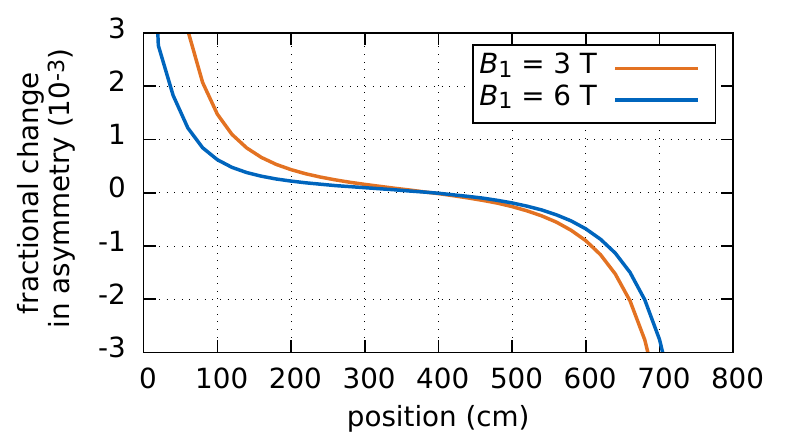}
	\caption{Fractional change of the measured asymmetry calculated for a point source located on the central axis at position $z$. The reference point is the centre of the long solenoid at $z=\unit{4}{m}$. (Field calculation without return yoke.)}
	\label{figure:B0 correct}
\end{figure}

In order to avoid decay particles staying within the decay region for extended periods of time, the field in the decay region is not completely uniform, but slightly increases towards the downstream end, see Fig.~\ref{figure:perc field0}. This inhomogeneity of $B_0$ within the decay volume influences the emission angle selection for the \ep. Thus the measured asymmetry changes according to Eq.~\eqref{form:asymmetry perc}. This effect is shown in Fig.~\ref{figure:B0 correct}. We note that the field uniformity perpendicular to the beam is much better, so only the direction along the beam axis counts.

For measurements with a continuous beam, this correction to the asymmetry would have to be integrated over the decay volume taking the neutron density into account. This would also include the strong contributions (and related uncertainties) at the ends of the long solenoid. A pulsed neutron beam can be used to minimise and control this effect as shown in Refs.~\cite{Mae09, Mae18} leading to a $10^{-4}$ effect only.

\subsection{Selector region}
\label{section:perc B1}

The purpose of the selector region is two-fold: it separates the \ep-beam from the neutron beam and provides the high field region for the phase space selection. The \emph{filter coils} allow to change the magnetic field strength from $3$ to $\unit{6}{T}$ in order to vary the filter effect. This split-pair design ensures a very good field uniformity over the entire field range. A pair of correction coils at both ends of the selector improves the uniformity of $B_1$ and eliminates minima between the selector and bending coils.
 
The \ep-beam is decoupled from the neutron beam using two tilted \emph{bending coils} and guided to the selector coil assembly, as shown in Fig.~\ref{figure:perc sketch2}.
The region from the end of the neutron guide to the point where the \ep-beam is separated from the neutrons is situated in an increasing magnetic field. This leads to large corrections with sizable uncertainties to an asymmetry measurement with a continuous beam. Thus the decoupling distance is designed to be as short as reasonably possible, while any local field minima must be avoided. Assuming an $m = 2$ neutron supermirror inside PERC, the neutron beam will have a divergence of $\approx\unit{2}{\degree}$ at a mean wavelength of $\lambda = \unit{0.5}{nm}$. The resulting decoupling distance then is $\unit{40}{cm}$.

\begin{figure}[tb]
	\centering
	\includegraphics[]{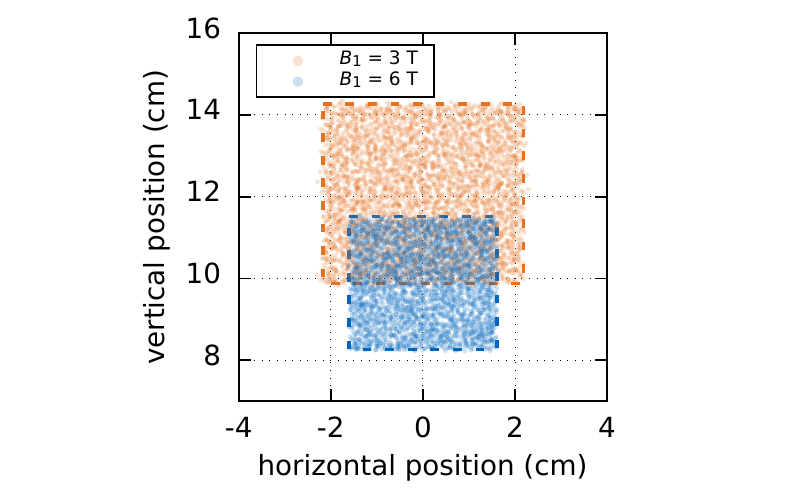}
	\caption{Simulated \ep-beam cross-section at the center plane of the selector coil, with $B_1 = \unit{3}{T}$ and $\unit{6}{T}$. Dots represent simulated electrons tracks, while the frames correspond to field lines from the edges of the neutron beam. {The vertical zero position} corresponds to the axis of the long solenoid.}
	\label{figure:e-monitor B1}
\end{figure}

In order to accommodate the neutron beamstop with its radiation shielding in the selector, the \textit{separation distance} between the \ep-beam and the neutron beam centers is required to be larger than $\unit{10}{cm}$. As a consequence of the asymmetry in the magnet layout and as a compromise to the complexity of the set-up, this beam separation is allowed to vary with a change in the $B_1$ field strength. Fig.~\ref{figure:e-monitor B1} shows the simulated \ep-beam distribution in the center plane of the selector for $B_1 = \unit{3}{T}$ and $\unit{6}{T}$. The \ep-beam shifts by as much as $\unit{2}{cm}$. We note that this requires the \ep-diaphragm, which is located there to limit edge effects, to be adjustable in size and position, see \cite{Dub08} for details.

\begin{figure}[!htb]
	\centering
	\includegraphics[]{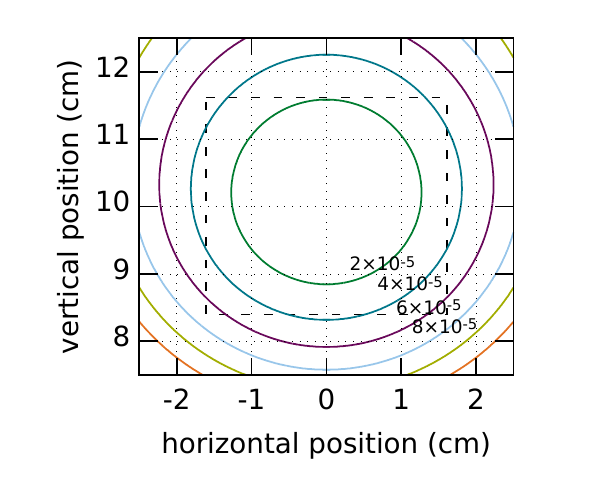}
	\caption{Contour plot of the relative change of the field $B_1 = \unit{6}{T}$ in the center plane of the selector. The black dashed frame indicates the cross-section of the \ep-beam.}
	\label{figure:B1 homo}
\end{figure}

In order to guarantee the required field uniformity in the center plane on the $10^{-4}$ level, the magnet design has to match this shift with $B_1$ in the \ep-beam trajectory. The point of the field minimum in this plane must shift accordingly using the additional \emph{filter coils}. Fig.~\ref{figure:B1 homo} shows the relative deviation of the $B_1$ field over the \ep-beam cross-section. The maximum effect is $1.3\times 10^{-4}$ for $B_1 = \unit{3}{T}$ and $7\times 10^{-5}$ for $B_1 = \unit{6}{T}$.

The \ep-transport has to fulfill the adiabatic condition $\gamma = \Delta B / B \ll 1$, with the field gradient $\Delta B$ over one helical pitch. Within the decay volume $\gamma \approx 0.003$ is negligibly small. Only in a small region at the end of the filter region $\gamma$ reaches a still tolerable value of $0.017$. At the rear end of PERC $\gamma$ increases due to the drop of the magnetic field. We note that detector systems are either placed inside the detector region or the adiabatic transport must be ensured by additional connecting magnet systems.

\section{Technical Aspects} %

A number of technical aspects influence the magnet design. For the production of PERC copper strands enclosing the superconducting wire with a rectangular cross section are used (wire-in-channel). Due to the helical winding scheme, the effective length of a solenoid is about one turn shorter than the coil former. Also the coils require enough space between them for the necessary mechanical supports. In the following we discuss some selected technical aspects.

\subsection{Quench safety}
\label{section:perc quench}

We use a superconducting wire based on Niobium Titanium alloy (NbTi). It consists of a bunch of twisted filaments with diameters of $\unit{78}{\micro m}$ or $\unit{42}{\micro m}$ embedded in a copper matrix. This round wire is soldered in a rectangular copper strand \cite{TDR}.

\begin{figure}[!htb]
	\centering
		\includegraphics[]{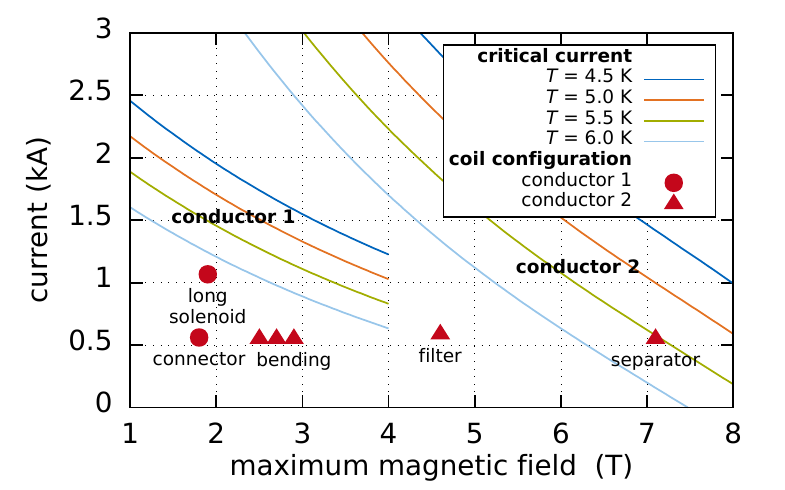}
	\caption{Critical current $I_\mathrm{c}$ versus magnetic field $B$ at different temperatures $T$, calculated according to \cite{Mai85} using the nominal wire specification of the two coductors used \cite{TDR,EAS15}. The markers denote the configuration of the different coils of PERC at $B_1 = \unit{6}{T}$.}
	\label{figure:quench diagram}
\end{figure}

The quenching of a superconductor wire is a function of the temperature $T$, the local magnetic field $B$, the current density $J$, and other factors like mechanical disturbances. Empirical formulae for the critical current density $J_\mathrm{c}(B,T)$ can for example be found in \cite{Lub83,Mai85,Bot00}.
The highest magnetic field at the wire of the magnet is $< \unit{7.2}{T}$ and occurs at the inner surface of the selector coil. Fig.~\ref{figure:quench diagram} shows the critical current $I_\mathrm{c}$ as a function of $B$ and $T$, and the configuration of the different PERC solenoids. We note that the actual measured critical current of the wires exceed the nominal specification by $10$ to $20\,\%$ \cite{EAS15}.

\subsection{Magnetic shielding}

In order to contain its high magnetic field, PERC will be equipped with a magnetic field return made of iron and steel. 
Four long iron bars are mounted alongside the magnet. At the high-field region, the field return is enforced by steel plates. As a result, the magnetic field strength decreases to the cardiac pacemaker limit of $\unit{0.5}{mT}$ within $\unit{5}{m}$ from the central magnet axis.

The non-linear material of the shielding also influences the magnetic field inside PERC. The maximal change of the field is 3.7\% at the ends of PERC. The field in the center of PERC changes by less than 0.5\%. These changes do not induce local field minima. The homogeneity in the \ep-beam cross-section in the selector coil is not influenced significantly \cite{Wang13dis}.

\section{Conclusion, Outlook}

With the design discussed in this paper the magnet system of PERC fulfils all criteria outlined in the instrument proposal which also includes an analysis of systematics \cite{Dub08}. In addition, the range of possible field configurations $B_0$, $B_1$, $B_2$ has been extended considerably, albeit at a lower nominal field due to superconductor constraints. 

The improved separator design allows for excellent shielding of the main detectors downstream from beam-related background. It also ensures stable particle trajectories and the possibility to couple secondary spectrometers to PERC.
The drawback of this scheme is that it is more difficult to monitor the neutron polarisation behind PERC since detector systems and beamstop have to be removed first. Using polarised \textsuperscript{3}He target cells, neutron polarisation can be measured at the $10^{-4}$ level required for PERC \cite{Klauser13,Klauser16}.

It is foreseen that PERC will be equipped with an active particle dump at the upstream end to detect particles reflected by the magnet filter or backscattered from the downstream detector \cite{Zie15}.  In addition, this enables a time-of-flight technique for detector calibration \cite{Roi18a}.

The technical design of the PERC magnet system is described in \cite{TDR}. The magnet system is currently in production and the delivery is expected in spring 2020.


\begin{acknowledgement}
The authors gratefully acknowledge the excellent support by the FRM~II and the engineers and workshops at the various institutions. We thank in particular K.~Lehmann (FRM), M. Horvath (ATI), and B.~Windelband and his team (HD). We thank F.~Glück for providing his magnetic field codes.
This work is supported by the Priority Programme SPP~1491 of German Research Foundation (DFG), contracts No. MA4944/1-1, MA4944/1-2, AB128/5-2 HE2308/9-1, SCH2708/1-1, SO1058/1-1, SO1058/2-1 and ZI816/1-1, the Austrian Science Fund (FWF), contracts No. I~528-N20, I~534-N20 and P~26636-N20, the University of Heidelberg, the FRM~II, and the DFG cluster of excellence \emph{Origin and Structure of the Universe}.
\end{acknowledgement}





\end{document}